\documentstyle[eqsecnum,prd,aps,preprint,tighten,epsf]{revtex}

\draft
\begin{document}

\def\beqa{\begin{eqnarray}}
\def\eeqa{\end{eqnarray}}
\def\p{\partial}
\def\lsim{\mathrel{\mathpalette\Oversim<}}
\def\gsim{\mathrel{\mathpalette\Oversim>}}
\def\Oversim#1#2{\lower0.5ex\vbox{\baselineskip0pt\lineskip0pt%
            \lineskiplimit0pt\ialign{%
          $\mathsurround0pt #1\hfil##\hfil$\crcr#2\crcr\sim\crcr}}}

\thispagestyle{empty}
{\baselineskip0pt
\leftline{\large\baselineskip16pt\sl\vbox to0pt{\hbox{\it Department of Physics}
               \hbox{\it Kyoto University}\vss}}
\rightline{\large\baselineskip16pt\rm\vbox to20pt{\hbox{KUNS 1398}
           \hbox{YITP-96-20}
               \hbox{\today}
\vss}}%
}
\vskip1cm
\begin{center}{\large \bf
Innermost Stable Circular Orbit of Coalescing}

{\large \bf Neutron Star-Black Hole Binary
}

\vspace{0.3cm}
{\large \bf
--- {\it Generalized Pseudo-Newtonian Potential Approach} ---
}
\end{center}
\vskip1cm
\begin{center}
 {\large
Keisuke Taniguchi {\footnote{
e-mail: taniguci@tap.scphys.kyoto-u.ac.jp}}
} \\
{\em Department of Physics,~Kyoto University,~Kyoto 606-01,~Japan}
\end{center}

\begin{center}
 {\large
Takashi Nakamura
} \\
{\em Yukawa Institute for Theoretical Physics,~Kyoto University,~Kyoto 606-01,~Japan}
\end{center}

\vskip1.5cm

\begin{center}
To appear in {\em Progress of Theoretical Physics, Vol.96 No.4}
\end{center}

\begin{abstract}

The innermost stable circular orbits (ISCO)
of coalescing neutron star-black hole binary are studied
taking into account both the tidal and relativistic effects.
We adopt the generalized pseudo-Newtonian potential
to mimic the general relativistic effects of gravitation.
It is found that the separation of the neutron star
and the black hole at the innermost stable circular orbit
is greater than that obtained
by using either the Newtonian potential
(i.e. the case in which only the tidal interaction is included.)
or the second post-Newtonian equation of motion of point mass
(i.e. the case in which the effect of general relativity
is taken into account but the tidal force is neglected.). 
In equal mass cases, it is found that for $\bar{a}/m_1 \gsim 3.5$
with $\bar{a}$ and $m_1$
being the mean radius and the mass of the neutron star,
the tidal effect dominates the stability of the binary system
while for $\bar{a}/m_1 \lsim 3.5$, the relativistic effect,
i.e. the fact that the interaction potential has unstable orbit, does.
The effect of the circulation is also studied
by comparing the Roche ellipsoids (REs)
with the irrotational Roche-Riemann ellipsoids (IRREs).
In the IRREs case, which are believed to be the case of
coalescing binary neutron stars,
it is found that in the equal mass binary case,
the orbital frequency is $495{\rm Hz}$ at the ISCO.

\end{abstract}

\section{INTRODUCTION}

The laser interferometers like LIGO\cite{LIGO}, VIRGO\cite{VIRGO},
GEO\cite{GEO} and TAMA\cite{TAMA} are currently constructed
and the detection of gravitational waves is expected
at the end of this century.
One of the most important sources of gravitational waves 
for these detectors is
coalescing binary compact stars such as
NS-NS, NS-BH and BH-BH binaries.
Each mass, each spin and the distance of the binary
can be determined by applying the matched filter techniques\cite{CF}
to the gravitational wave form
of the so-called last three minutes of the binary\cite{C}.
In the end of the last three minutes two compact stars coalesce
and nonlinear character of the gravity
and the tidal effects become important,
which will be the most exciting part of the coalescing event.

It is known from the study of orbits of a test particle
in the Schwarzschild metric
that the innermost stable circular orbit (ISCO) exists at $r=6M$.
For binary cases, 
Kidder, Will, \& Wiseman\cite{KWW} investigated a point-mass binary
using second post-Newtonian equation of motion,
and found that the ISCO of the comparable mass binary
is at $r_{gr} \simeq 7M_{tot}$
where $r_{gr}$ is the separation of the binary
in the Schwarzschild like radial coordinate.

As for the tidal effects, Chandrasekhar\cite{Ch69} studied
the Roche limit by using Newtonian gravity and
treating binary systems as incompressible homogeneous ellipsoids.
He found that the stars are tidally disrupted before contact
at $r_t=(2.25M_{tot}/\rho)^{1/3}$ for equal mass binary
where $\rho$ is the constant density of ellipsoids.
For the typical neutron star with $M_{tot}=2.8M_\odot$ and
$\rho =1 \times 10^{15}{\rm g/cm^3}$, $r_t$ becomes $5.6M_{tot}$.
 
Recently Lai, Rasio, \& Shapiro\cite{LRS1}\cite{LRS2} have
investigated the binary consist of finite size compressible stars
using  approximate equilibria.
In a series of their papers, they took into account
the effect of the quadrupole order deviation of stars 
and found that the hydrodynamic instability occurs
at $r_h=6\sim 7.5 M_{tot}$ for $n=0.5$ polytropic neutron star
with the radius $R_0=2.5M_{tot}$.

In order to know at what radius the final merging phase begins, 
all three effects, that is,
the general relativistic, tidal and hydrodynamic effects
should be taken into account simultaneously
since $r_{gr}, r_t$ and $r_h$ have similar values of $\sim 7M_{tot}$.
For this purpose 
we must solve the fully general relativistic equations,
which is a difficult 3D problem in numerical relativity\cite{N}
although the partly general relativistic results
have already been presented \cite{WM} \cite{Shibata}.
To know the qualitative feature of the ISCOs
we will use various approximations
to calculate equilibria of the binary in this paper.
The reward is that all the calculations can be done analytically
so that our results will contribute some physical aspects
to the final understanding of the ISCO.

In this paper, we will solve the Roche problem\cite{Ch69}
as a model of the binary
that consists of a finite size star and a point-like gravity source.
A gravitational potential which has unstable circular orbits
will be used as an interaction potential of the binary.
Specifically, we generalize the pseudo-Newtonian potential
proposed by Paczy$\acute{{\rm n}}$sky \& Wiita\cite{PW}
to mimic the general relativistic effects.
We think that this model mimics a neutron star-black hole binary,
and describes its qualitative behavior.

This paper is organized as follows.
In \S2 the basic equations necessary for constructing
equilibrium configurations of the Roche ellipsoids (REs)
and the Roche-Riemann ellipsoids (RREs)\cite{Aizenman}
are derived in the case that the interaction potential is general.
In \S3 circular orbits of the neutron star-black hole binary
are calculated using the generalized pseudo-Newtonian potential
to mimic the effects of general relativity
and the results are shown in \S4.
In \S5 our results are compared
with those using the Newtonian potential
and the second order post-Newtonian equation of motion .
We use the units of $c=G=1$ through this paper.

\section{GENERALIZED ROCHE-RIEMANN ELLIPSOIDS}

We first regard the black hole as a point particle of mass $m_2$
and denote the gravitational potential by the black hole as $V_2(r)$.
To mimic the effects of general relativity by $V_2(r)$,
we do not fix the form of $V_2(r)$ at the moment.   
Next we treat the neutron star
as an incompressible, homogeneous ellipsoid
of the semi-axes $a_1, a_2$ and $a_3$,
mass $m_1$ and the density $\rho_1$.
We treat the self-gravity of the neutron star ($V_1$) as Newtonian.
Although for mathematical convenience
we assume the incompressibility for the equation of state, 
the effect of the compressibility can be easily taken into account
in an approximate way\cite{LRS1}.
Following Chandrasekhar\cite{Ch69},
we call the neutron star the primary and the black hole the secondary.

\subsection{Second Virial Equations}

We use the tensor virial method\cite{Ch69},
and derive the equations necessary
for constructing equilibrium figures of this system.
Choose a coordinate system such that
the origin is at the center of mass of the primary,
the $x_1$-axis points to the center of mass of the secondary,
and $x_3$-axis coincides with the direction of 
the angular velocity of the binary ${\bf \Omega}$. 
In the frame of reference rotating with ${\bf \Omega}$,
the Euler equations of the primary are written as
\beqa
        \rho_1 \frac{du_{i}}{dt} = -\frac{\p P}{\p x_{i}} +
        \rho_1 \frac{\p}{\p x_{i}} \left[  V_{1} + V_{2} +
        \frac{1}{2} {\Omega}^2 \left\{ {\left(
        \frac{m_{2} R}{m_{1}+m_{2}}
        - x_1 \right)}^2 + x_2^2 \right\} \right] +
        2 \rho_1 \Omega \epsilon_{il3} u_{l},  \label{eom}
\eeqa
where $\rho_1, u_i, P$ and $R$ are the density, the internal velocity,
the pressure and the separation
between the neutron star and the black hole, respectively.

Let us expand the interaction potential $V_2$
in power series of $x_k$ up to the second order.
This approximation is justified
if $R$ is much larger than $a_1, a_2$ and $a_3$. 
We assume that the potential $V_2$ is spherically symmetric
so that it depends only on the distance $r$
from the center of mass of the secondary as
\beqa
        V_2 = V_2(r),
\eeqa
where $r$ is given by
\beqa
        r=\left\{(R-x_1)^2 +x_2^2 +x_3^2\right\}^{1/2}.
\eeqa
The expansion of  $ V_2(r)$ becomes
\beqa
        V_2 = (V_2)_0 - \left( \frac{\p V_2}{\p r}
        \right)_0 x_1 + \frac{1}{2} \left(
        \frac{\p^2 V_2}{\p r^2} \right)_0 x_1^2 +
        \frac{1}{2R} \left( \frac{\p V_2}{\p r}
        \right)_0 \left( x_2^2 + x_3^2 \right), \label{V2}
\eeqa
where the subscript $0$ denotes the derivatives 
at the origin of the coordinates.
In the case of the circular orbit, we have
from the force balance at the center
\beqa
        \frac{m_2 R}{m_1+m_2} {\Omega}^2 = -{\left( \frac{\p V_2}
        {\p r} \right)}_0 (1+\delta),  \label{omefo}
\eeqa
where $\delta$ is the quadrupole term
of the interaction potential\cite{LRS1}.

Substituting Eqs.(\ref{V2}) and (\ref{omefo}) into Eq.(\ref{eom}),
we have
\beqa
        \rho_1 \frac{du_{i}}{dt} = -\frac{\p P}{\p x_{i}}
        &+& \rho_1 \frac{\p}{\p x_{i}} \left[  V_{1} + \delta
        \left( \frac{\p V_2}{\p r} \right)_0 x_1+
        \frac{1}{2} {\Omega}^2 \left( x_1^2 + x_2^2 \right) +
        \frac{1}{2} \left( \frac{{\p}^2 V_2}
        {\p r^2} \right)_0 x_1^2  \right. \nonumber \\
        &+& \left. \frac{1}{2R} \left( \frac{\p V_2}{\p r}
        \right)_0 \left( x_2^2 + x_3^2 \right) \right]
        + 2 \rho_1 \Omega \epsilon_{il3} u_l.  \label{feom}
\eeqa
Multiplying $x_j$ to Eq.(\ref{feom}) 
and integrating over the volume of the primary,
we have
\beqa
        \frac{d}{dt} \int \rho_1 u_i x_j d^3 {\bf x} &=&
        2 T_{ij} + W_{ij} + \left\{ {\Omega}^2 + \left(
        \frac{{\p}^2 V_2}{\p r^2} \right)_0 \right\}
        \delta_{1i} I_{1j}  \nonumber  \\
        & & + \left\{  {\Omega}^2 + \frac{1}{R}
        \left( \frac{{\p} V_2}{\partial r} \right)_0 \right\}
        \delta_{2i} I_{2j}
        + \frac{1}{R} \left( \frac{{\p} V_2}
        {\partial r} \right)_0 \delta_{3i} I_{3j}  \nonumber  \\
        & & + 2 \Omega \epsilon_{il3} \int \rho_1 u_l x_j
        d^3 {\bf x} + \delta_{ij} \Pi,   \label{sve}
\eeqa
where
\beqa
        T_{ij} &\equiv& \frac{1}{2} \int \rho_1 u_{i} u_{j} d^3 {\bf x} :
                {\rm Kinetic~Energy~Tensor},   \\
        W_{ij} &\equiv& \int \rho_1 \frac{\partial V_1}{\partial x_i}
                x_{j} d^3 {\bf x} : {\rm Potential~Energy~Tensor},  \\
        I_{ij} &\equiv& \int \rho_1 x_{i} x_{j} d^3 {\bf x} ~~~:
                {\rm Moment~of~Inertia~Tensor},
\eeqa
and
\beqa
        \Pi &\equiv& \int P d^3 {\bf x}.
\eeqa
In Eq.(\ref{sve}) there is no terms related to $\delta$.
Since it is possible to take the coordinate system comoving
with the center of mass of the binary system,
the term proportional to $\delta$ in Eq.(\ref{feom}) vanishes
when we integrate over the volume of the primary.
Eq.(\ref{sve}) is the basic equation
to construct the equilibrium figures of the Roche ellipsoids (REs)
and the Roche-Riemann ellipsoids (RREs) 
for the general potential $V_2(r)$.

\subsection{Equilibrium Roche-Riemann Sequence}

In this subsection,
we show how to construct the equilibrium figures of the RREs.
The Roche-Riemann ellipsoid is the equilibrium
in which the shape of the primary does not change in the rotating frame
although the uniform vorticity exists inside the primary. 
We restrict the problem to the simplest case
where the uniform vorticity of the primary
is parallel to the rotation axis, i.e.
the primary is the Riemann S-type ellipsoid\cite{Ch69}.

We set the coordinate axes
to coincide with the principal axes of the primary.
For the uniform vorticity $\zeta$,
the internal velocity $u_i$ in the rotating frame is given by
\beqa
        u_1 &=& Q_1 x_2,  \\
        u_2 &=& Q_2 x_1,  
\eeqa
and
\beqa
        u_3 &=& 0,
\eeqa
where
\beqa
        Q_1 &=& - \frac{a_1^2}{a_1^2 + a_2^2} \zeta
\eeqa
and
\beqa
        Q_2 &=& \frac{a_2^2}{a_1^2 + a_2^2} \zeta.
\eeqa
For the stationary equilibrium, Eq.(\ref{sve}) is rewritten as
\beqa
        Q_{ik}Q_{jl}I_{kl} + W_{ij} &+& \left\{ {\Omega}^2 + \left(
        \frac{{\p}^2 V_2}{\p r^2} \right)_0
        \right\} \delta_{1i} I_{1j} + \left\{  {\Omega}^2
        + \frac{1}{R} \left( \frac{{\p} V_2}{\p r}
        \right)_0 \right\} \delta_{2i} I_{2j} \nonumber  \\
        &+& \frac{1}{R} \left( \frac{{\p} V_2}{\p r}
        \right)_0 \delta_{3i} I_{3j} + 2 \Omega \epsilon_{il3}
        Q_{lk} I_{kj} = - \delta_{ij} \Pi,  \label{stasve}
\eeqa
where $Q_{ij}$ is not zero only for
\beqa
        Q_{12} &=& Q_1, \\
        Q_{21} &=& Q_2. 
\eeqa
Eq.(\ref{stasve}) has only diagonal components  as
\beqa
        Q_1^2 I_{22} + W_{11} + \left\{ {\Omega}^2 +
        \left( \frac{{\p}^2 V_2}
        {\p r^2} \right)_0 \right\} I_{11} +
        2 \Omega Q_2 I_{11} &=& - \Pi, \label{xsve}  \\
        Q_2^2 I_{11} + W_{22} + \left\{ {\Omega}^2 + \frac{1}{R}
        \left( \frac{{\p} V_2}
        {\p r} \right)_0 \right\} I_{22} -
        2 \Omega Q_1 I_{22} &=& - \Pi, \label{ysve}
\eeqa
and
\beqa
        W_{33} + \frac{1}{R} \left( \frac{{\p} V_2}
        {\p r} \right)_0 I_{33} &=& - \Pi.  \label{zsve}
\eeqa

We assume for simplicity that the gravitational potential
of the primary is Newtonian.
In this case, the potential energy tensor
and the moment of inertia tensor
of the incompressible, homogeneous ellipsoids are calculated as
\beqa
        W_{ij} &=& -2 \pi \rho_1 A_{i} I_{ij},
\eeqa
and
\beqa
        I_{ij} &=& \frac{1}{5} m_1 a_{i}^2 \delta_{ij},
\eeqa
where
\beqa
        A_{i} &=& a_{1} a_{2} a_{3} \int_0^{\infty} \frac{du}
        {\Delta \left(a_{i}^2 +u \right)} , 
\eeqa
and
\beqa
        \Delta^2 &=&  \left( a_{1}^2 +u \right)
        \left( a_{2}^2 +u \right) \left( a_{3}^2 +u \right).  
\eeqa

Eliminating $\Pi$ from Eqs.(\ref{xsve})-(\ref{zsve}), 
we have
\beqa
        \left[ \left\{ 1+ 2\frac{a_2^2}{a_1^2+a_2^2} f_R +
        \left( \frac{a_1a_2}{a_1^2+a_2^2} f_R \right)^2 \right\}
        \Omega^2 + \left( \frac{\p^2 V_2}{\p r^2}
        \right)_0 \right] a_{1}^2 &-& \frac{1}{R} \left(
        \frac{{\p} V_2}{\p r} \right)_0 a_{3}^2 \nonumber \\
        &=& 2 \pi \rho_1
        \left( a_{1}^2 - a_{3}^2 \right) B_{13}  \label{sve1}
\eeqa
and
\beqa
        \left[ \left\{ 1+ 2\frac{a_1^2}{a_1^2+a_2^2} f_R +
        \left( \frac{a_1a_2}{a_1^2+a_2^2} f_R \right)^2 \right\}
        \Omega^2 + \frac{1}{R} \left( \frac{\p V_2}{\p r} \right)_0
        \right] a_{2}^2 &-& \frac{1}{R} \left( \frac{\p V_2}
        {\p r} \right)_0 a_{3}^2 \nonumber \\
        &=& 2 \pi \rho_1 \left( a_{2}^2 - a_{3}^2 \right) B_{23},
        \label{sve2}
\eeqa
where
\beqa
        f_{R} \equiv \frac{\zeta}{\Omega}
\eeqa
and the following relations are used;
\beqa
        a_i^2 A_i -a_j^2 A_j &=&
        \left( a_i^2-a_j^2 \right) B_{ij},
\eeqa
where
\beqa
        B_{ij} &=&a_1 a_2 a_3 \int_0^{\infty} \frac{u du}
        {\Delta (a_i^2 +u)(a_j^2 +u)}.
\eeqa
Now from Eq.(2.5) $\Omega$ is given by
\beqa
        \Omega^2 = -\frac{1+p}{R} \left( \frac{\p V_2}{\p r}
        \right)_0 (1+\delta)
        ~~~~~~~~~~\left( p \equiv \frac{m_1}{m_2} \right).
        \label{omega}
\eeqa

Dividing Eq.(\ref{sve1}) by Eq.(\ref{sve2}),
we have the equation to determine the Roche-Riemann sequences as
\beqa
        \frac{\left[ (1+p)(1+\delta)
        \left\{ 1+ 2\frac{a_2^2}{a_1^2+a_2^2} f_R
        +\left( \frac{a_1 a_2}{a_1^2+a_2^2} f_R \right)^2 \right\}
        -R \left( \frac{\p^2 V_2}{\p r^2} \right)_0/
        \left( \frac{\p V_2}{\p r} \right)_0 \right]a_1^2 +a_3^2}
        {\left[ (1+p)(1+\delta)
        \left\{ 1+2 \frac{a_1^2}{a_1^2+a_2^2} f_R
        +\left( \frac{a_1a_2}{a_1^2+a_2^2} f_R \right)^2 \right\}
        -1 \right] a_2^2 + a_3^2}
        = \frac{\left( a_1^2 - a_3^2 \right) B_{13}}
        {\left( a_2^2 - a_3^2 \right) B_{23}}. \label{gaxis}
\eeqa
Using Eqs.(\ref{sve2}) and (\ref{omega}),
we can determine the orbital angular velocity $\Omega$ by
\beqa
        \frac{\Omega^2}{\pi \rho_1}=\frac{2(1+p)(1+\delta)\left(a_2^2
        -a_3^2 \right) B_{23}}{ \left[ (1+p)(1+\delta)\left\{ 1+
        2 \frac{a_1^2}{a_1^2+a_2^2} f_R + \left(
        \frac{a_1a_2}{a_1^2+a_2^2} f_R \right)^2 \right\}-1
        \right] a_2^2 + a_3^2}. \label{gOme}
\eeqa
Note that $f_R$ is related to the circulation ${\cal C}$ as
\beqa
        {\cal C} &=& \oint {\bf u}_{inertial} \cdot d{\bf l}=
\pi a_1 a_2 (2 + f_R) \Omega, 
\eeqa
where
\beqa
        {\bf u}_{inertial} &=& \left\{
             \begin{array}{@{\,}ll}
              (u_{inertial})_1 = (Q_1 -\Omega) x_2, \\
              (u_{inertial})_2 = (Q_2 + \Omega) x_1-\frac{R}{1+p}
                \Omega~ \\
              (u_{inertial})_3 = 0.
             \end{array}
            \right.
\eeqa
If there is no viscosity inside the primary,
the circulation should be conserved from Kelvin's circulation theorem.

\subsection{Total Angular Momentum}

The total energy and the total angular momentum of the binary
are the decreasing functions of time
since the gravitational waves are emitted.
If the total angular momentum has its minimum at some separation
of the binary, we regard this point as the ISCO
\footnote{Lai, Rasio, \& Shapiro show in appendix D of \cite{LRS1}
that the true minimum point of the total energy coincides with
that of the total angular momentum.
Strictly speaking, 
if the rotation includes only to the quadrupole order,
this coincidence fails.
However the difference is as small as the numerical accuracy
\cite{LRS1}.}.
The total angular momentum of our system
which is the sum of the orbital and the spin angular momentum
is given by
\beqa
        J_{tot} &=& m_1 r_{cm}^2 \Omega + m_2 (R-r_{cm})^2 \Omega +
        I \Omega + \frac{2}{5}m_1 \frac{a_1^2 a_2^2}
        {a_1^2 + a_2^2} \zeta \nonumber  \\
        &=& \frac{m_1 m_2}{m_1 + m_2} R^2 \Omega
        \left\{ 1 + \frac{1}{5} (1+p) \frac{1}{R^2}
        \left( a_1^2 + a_2^2 + 2\frac{a_1^2 a_2^2}{a_1^2+a_2^2}
        f_R \right) \right\}   \label{angmom}
\eeqa
where
\beqa
        r_{cm} = \frac{m_{2} R}{m_1+m_2}.
\eeqa
The first term in the braces
of the right hand side of Eq.(\ref{angmom}) comes
from the orbital angular momentum of the binary system
and the second does from the spin angular momentum of the primary.

\section{GENERALIZED PSEUDO-NEWTONIAN POTENTIAL}

There are variety of choices of $V_2(r)$
to mimic the general relativistic effects of the gravitation.
We generalize the so-called pseudo-Newtonian potential
proposed by Paczy$\acute{{\rm n}}$sky \& Wiita\cite{PW} originally. 
This potential fits the effective potential
of the Schwarzschild black hole quite well as we will show later.
We will use the generalized pseudo-Newtonian potential defined by
\beqa
        V_2(r) &=& \frac{m_2}{r-r_{pseudo}} \label{pnpot},\\
        r_{pseudo} &=& r_s \left\{ 1+ g(p) \right\}, \\
        g(p) &=& \frac{7.49p}{6(1+p)^2} - \frac{10.4 p^2}{3(1+p)^4}
        + \frac{29.3 p^3}{6(1+p)^6},  \label{gp} \\
        r_s &\equiv& \frac{2GM_{tot}}{c^2}, \\
        M_{tot} &=& m_1 + m_2, 
\eeqa
where $p=m_1/m_2$ and $g(p)$ is the special term
to fit the ISCOs of the hybrid second post-Newtonian calculations
by Kidder, Will, \& Wiseman\cite{KWW}.
For $p=0$, the generalized pseudo-Newtonian potential
agrees with the pseudo-Newtonian potential
proposed by Paczy$\acute{{\rm n}}$sky \& Wiita\cite{PW}.

Fig.1(a) shows effective potentials (solid lines) and 
locations of circular orbits (dots)
in our generalized pseudo-Newtonian potential
( $p=0$ \& $r_{pseudo}=r_s$) and in the Schwarzschild metric. 
Although by this choice of the parameter ($r_{pseudo}=r_s$), 
the locations of the ISCOs in the generalized pseudo-Newtonian potential
agree with those in the Schwarzschild metric,
the angular momenta at the ISCO are different,
that is, the angular momentum
in the generalized pseudo-Newtonian potential ($J_{pseudo}$)
for $p=0$ is $(9/8)^{1/2}$ times larger than
that in the Schwarzschild metric ($J_{Sch}$) at the ISCO. 
Therefore in Fig.1(a) and (b) we compare circular orbits with
different angular momentum related as
\beqa
        J_{pseudo} = \left( \frac{9}{8} \right)^{1/2} J_{Sch}.
\eeqa
   From Fig.1(b) we see that the radii of the circular orbits of 
the generalized pseudo-Newtonian potential agrees with
those of the effective potential
around Schwarzschild black hole within 10\% accuracy near the ISCO.
This is the reason why we believe that
our generalized pseudo-Newtonian potential
expresses the effect of general relativity within 10\% or so.

Using Eq.(\ref{pnpot}) and Eq.(\ref{omega}),
we can rewrite Eq.(\ref{gOme}) as
\beqa
        \frac{p^2 r_s^3 (\bar{a}/m_1)^3}{12(1+p)^3 R(R-r_{pseudo})^2}
        - \frac{(a_2^2-a_3^2) B_{23}}
        { \left[ (1+p)(1+\delta)
        \left\{ 1+ 2 \frac{a_1^2}{a_1^2 + a_2^2} f_R +
        \left( \frac{a_1a_2}{a_1^2 + a_2^2} f_R \right)^2 \right\}
        -1 \right] a_2^2 + a_3^2 } = 0, \label{sequence}
\eeqa
where $\bar{a}$ is the mean radius of the primary.
In the generalized pseudo-Newtonian case,
the quadrupole term $\delta$ is written as
\beqa
        \delta=\frac{3}{10} \left\{2a_1^2-
        \frac{(3R-r_{pseudo})(R-r_{pseudo})}{3R^2}
        \left(a_2^2+a_3^2 \right) \right\}
        \frac{1}{(R-r_{pseudo})^2}. \label{quadru}
\eeqa
We also have the separation of the binary as
\beqa
        R=\frac{E}{E-2} r_{pseudo}, \label{sob}
\eeqa
where
\beqa
        E \equiv &-&(1+p)(1+\delta)
        \left\{ 1+ 2\frac{a_2^2}{a_1^2+a_2^2} f_R +
        \left( \frac{a_1 a_2}{a_1^2 + a_2^2} f_R \right)^2 \right\}
        - \frac{a_3^2}{a_1^2} \nonumber \\
        &+& \left\{ \left[ (1+p)(1+\delta) \left\{ 1 +
        2\frac{a_1^2}{a_1^2+a_2^2} f_R
        + \left( \frac{a_1a_2}{a_1^2+a_2^2} f_R \right)^2 \right\}
        -1 \right] \frac{a_2^2}{a_1^2} +
        \frac{a_3^2}{a_1^2} \right\}
        \frac{(a_1^2-a_3^2) B_{13}}{(a_2^2-a_3^2) B_{23}}.
\eeqa

For the given mass ratio $p$, mean radius $/bar{a}/m_1$,
circulation parameter $f_{R}$ and axial ratio $a_3/a_1$,
we can determine
the axial ratio $a_2/a_1$ by solving Eq.(\ref{sequence})
with Eqs.(\ref{quadru}) and (\ref{sob}).
Using the axial ratios ($a_2/a_1$, $a_3/a_1$),
we are able to calculate the orbital angular velocity by Eq.(\ref{gOme})
and the separation of the binary by Eq.(\ref{sob}).
The total angular momentum is calculated by Eq.(\ref{angmom}).
Finding the minimum of the total angular momentum,
we can determine the location of the ISCO.

When the viscosity inside the primary is so effective
that no internal motion exists,
$f_R=0$ in the above equation, i.e. the Roche ellipsoids (REs).
While if the primary is in-viscid and ${\cal C}=0$,
we have $f_R=-2$, i.e.
the irrotational Roche-Riemann ellipsoids (IRREs).
Note that if we substitute the Newtonian potential
as an interaction potential,
Eqs.(\ref{gaxis}) and (\ref{gOme}) agree with
the equations derived by Chandrasekhar\cite{Ch69}
in the REs ($f_R=0$) case and those by Aizenman\cite{Aizenman}
in the RREs case.

\section{RESULTS}

Kochanek\cite{Kocha} and Bildsten \& Cutler\cite{BC} showed
that the internal structure of a coalescing binary neutron star is
the irrotational Roche-Riemann ellipsoids (IRREs).
However we calculate both the REs and the IRREs for comparison.
We show the results in four cases;
1) the REs with  $p=1$ or 0.1,
2) the IRREs with  $p=1$ or 0.1. 
Results are given in Table.I to Table.VI,
where $\tilde{\Omega}=\Omega/\sqrt{\pi \rho_1}$ represents
the normalized orbital angular velocity,
and $\tilde{J}$ denotes the normalized total angular momentum
defined by
\beqa
        \tilde{J}=\frac{J_{tot}}{m_1 m_2 (r_s/M_{tot})^{1/2}}.
\eeqa
$\bar{a}$ is the mean radius of the primary defined by
\beqa
        \bar{a} = \left( \frac{m_1}{\frac{4}{3} \pi \rho_1}
        \right)^{1/3}.
\eeqa

In Tables I to VI, $\dagger$ means the point of the ISCO
defined in this paper,
and $\ddagger$ does the point of the Roche limit where
the Roche limit is defined by the distance of closest approach
for equilibrium to be possible\cite{Ch69}.
The values in the parentheses show the power of $10$.

\subsection{$p=1$ Case}

Fig.2(a) shows $\tilde{J}$
as a function of the normalized separation $R/r_s$.
In Fig.2(a), thin solid, dotted and dashed lines are
the REs with $\bar{a}/m_1$ being $3$, $5$ and $8$, respectively,
while thick solid, dotted and dashed lines are
the IRREs with $\bar{a}/m_1$ being $3$, $5$ and $8$, respectively.
We defined in \S\S II.C that the location of the ISCO is
the minimum point of $\tilde{J}$.
   From Fig.2(a), we see that the separations of the binary at the ISCO 
in the IRREs case are almost the same as those in the REs case.

Fig.2(b) shows the axial ratios $a_2/a_1$ and $a_3/a_1$
as a function of $R/r_s$.
The conventions of lines are the same as those in Fig.2(a).
The relation among the length of the axes is $a_1>a_2>a_3$ in the REs,
while $a_1>a_3>a_2$ in the IRREs.
In the REs, the tidal force makes the $a_1$ axis long
while it does $a_2$ and $a_3$ axes short.
The centrifugal force makes $a_1$ and $a_2$ axes long.
As a result we have $a_1>a_2>a_3$.
In the IRREs, in addition to the above effects,
the Coriolis force caused
by the internal motion of the primary exists.
This makes $a_1$ and $a_2$ axes short, which yields $a_1>a_3>a_2$ .

   From Fig.2(b), it is found that in the IRREs the star
with $\bar{a}/m_1=3$ reaches the ISCO
at the point of $a_2/a_1 \simeq 0.912$ and $a_3/a_1 \simeq 0.918$.
On the other hand, the star with $\bar{a}/m_1=8$ terminates
when $a_2/a_1 \simeq 0.760$ and $a_3/a_1 \simeq 0.792$.
This means that
the primary with the smaller mean radius reaches the ISCO
before the shape of the primary deviates from the sphere considerably.
This tendency is the same for the REs.

\subsection{ $p=0.1$ Case}

Fig.3(a) shows $\tilde{J}$ as a function of $R/r_s$.
The conventions are the same as those in Fig.2(a)
except $\bar{a}/m_1$ being $3$, $5$ and $8$.
   From Fig.3(a), we see that all lines with
the mean radii in the range of $3 \le \bar{a}/m_1 \le 8$
take their minimum at the points near $R/r_s \sim 3.25$.
This value is almost the same as the ISCO
by Kidder, Will, \& Wiseman\cite{KWW} for $p=0.1$. 
This means that when $p$ is much less than $1$ and
the mean radius of the primary is less than $8m_1$ which corresponds 
to 17km for $m_1=1.4M_\odot$,
the size of the primary has little effect on the ISCO.

Fig.3(b) shows the axial ratios $a_2/a_1$ and $a_3/a_1$
as a function of $R/r_s$.
The conventions are the same as those in Fig.3(a).
We see that the stars with smaller mean radii reach
the ISCO even when the deviation
from the spherical symmetry is very small.
Since the binary system enters an unstable circular orbit
before the primary is tidally deformed,
the tidal effects are not important.
We see that even $\bar{a}/m_1\sim 8$
the minimum values $a_2/a_1$ and $a_3/a_1$ are not small ($\sim 0.85$).

\section{DISCUSSIONS}

In this section, we discuss the differences between
the case of the generalized pseudo-Newtonian potential and
that of the Newtonian potential,
and between the REs and the IRREs.
We will also compare our results with other papers.

\subsection{The Case of $p=1$}
In Fig.4, the separation $R_{ISCO}/r_s$ is shown
as a function of $\bar{a}/m_1$.
Thick lines and thin lines represent
the case of the generalized pseudo-Newtonian potential and
that of the Newtonian potential, respectively.
Solid lines and dotted lines express the case of 
the IRREs and the REs, respectively.
We see that for the Newtonian potential,
$R_{ISCO}/r_s$ increases in proportion to $\bar{a}/m_1$
regardless of types of ellipsoids,
while for the generalized pseudo-Newtonian potential,
the behavior of $R_{ISCO}/r_s$ changes around
$\bar{a}/m_1 \simeq 3.5$.

For $\bar{a}/m_1 \gsim 3.5$,
$R_{ISCO}/r_s$ increases in proportion to $\bar{a}/m_1$ as
in the case of the Newtonian potential.
In this region, the tidal effect dominates the system and
the effects of the general relativity become less important.
In conclusion, for $\bar{a}/m_1 \gsim 3.5$,
the tidal effect dominates the stability of the binary system and
for $\bar{a}/m_1 \lsim 3.5$, the binary system is dominated
by the relativistic effect, i.e.
the fact that the interaction potential has an unstable orbit.
  From Fig.4 it is also found that the location of the ISCOs
in the IRREs case is not so different from that in the REs case
for the same $\bar{a}/m_1$.

The orbital frequency of the IRREs at the ISCO 
for $m_1=1.4M_{\odot}$ and $\bar{a}/m_1 = 5$ is estimated from
$\tilde{\Omega}$ in Table.IV as
\beqa
        \Omega_{\rm ISCO} &=& 495~[{\rm Hz}].
\eeqa
This value is smaller than that in the Newtonian potential case
($\Omega_{\rm ISCO}^{Newton} = 599~[{\rm Hz}]$).

\subsection{The Case of $p=0.1$}

If we consider the primary
as the neutron star of mass $1.4M_{\odot}$,
then, the secondary is regarded
as the black hole of mass $14M_{\odot}$.
$R_{ISCO}/r_s$ is shown
as a function of $\bar{a}/m_1$ in Fig.5.
For the Newtonian potential,
$R_{ISCO}/r_s$ increases
in proportion to $\bar{a}/m_1$ like the case of $p=1$.
For the generalized pseudo-Newtonian potential,
in the range of $\bar{a}/m_1$ of Fig.5,
$R_{ISCO}/r_s$ converges to the value $3.25$ obtained
by Kidder, Will, \& Wiseman\cite{KWW}.
This is because the existence of the unstable orbit
in the generalized pseudo-Newtonian potential
influences $R_{ISCO}/r_s$,
and this effect dominates
when the radius of the primary is small.
This is clearer for the small mass ratio $p$.
Therefore if $p$ is small, for the range of the radius
($3 \le \bar{a}/m_1 \le 8$)
relevant to the neutron star,
the effect of the neutron star's size is very small.
This comes essentially from the fact
that the Newtonian estimate of the Roche radius is
smaller than the radius of the ISCO.

One can estimate the orbital frequency of the IRREs
at the ISCO for $m_1=1.4M_{\odot}$
and $\bar{a}/m_1 = 5$ from $\tilde{\Omega}$ in Table.V as  
\beqa
        \Omega_{\rm ISCO} &=& 187~[{\rm Hz}].
\eeqa

\subsection{Comparison with Other Works}

Kidder, Will, \& Wiseman\cite{KWW} studied
the motion of the point-particle binary systems using
{\it hybrid Schwarzschild second post-Newtonian equation of motion},
and obtained the separation at the ISCO ($r_{ISCO}$) expressed by
\beqa
        \frac{r_{ISCO}}{M_{tot}} \simeq 6 + 7.49 \eta - 20.8 \eta^2
        +29.3 \eta^3  \label{KWWISCO}
\eeqa
where
\beqa
        \eta &=& \frac{m_1 m_2}{M_{tot}^2}. 
\eeqa

Table.VII shows the comparison of Eq.(\ref{KWWISCO})
with our results for $\bar{a}/m_1 = 5$.
It is found that for $p=0.1$, the finite size effect is not important
because the primary is much lighter than the secondary,
so that the results are almost the same as \cite{KWW}.
On the other hand, when $p=1$ the finite size of the primary
increases the separation at the ISCO
by the general relativistic and tidal effects.

Lai, Rasio, \& Shapiro\cite{LRS3},
discussed the relativistic effects on the binary system
for compressible Darwin ellipsoids
using a simple approximate model\cite{LRS4}.
The Darwin ellipsoids are the equilibrium configurations
constructed by two identical synchronized finite-size stars 
including the mutual tidal interactions.
In their approach the effects of general relativity
and the Newtonian tidal interactions
for finite-size compressible stars are combined by hand.
While we formulated the problem
using arbitrary interaction potentials of the secondary
for the incompressible primary. 
We adopted the semi-relativistic potential called 
the generalized pseudo-Newtonian potential
to mimic the general relativistic effects of gravitation.
We solved the equilibria of the REs and the IRREs in this potential.
Their results and ours are compared in Table.VIII,
where $r_m$ expresses the minimum separation obtained by
Lai, Rasio, \& Shapiro\cite{LRS3}.
   From this table, we see that both results agree
rather well inspite of different approaches and approximations.

\acknowledgments
KT would like to thank H. Sato, K. Nakao and M. Shibata
for useful discussions and continuous encouragement.
This work was in part supported by a Grant-in-Aid for Basic Research
of Ministry of Education, Culture, Science and Sports (08NP0801).

\newpage

\begin{center}
  {\large TABLE CAPTIONS}
\end{center}

\vspace{0.5cm}

Table.I. Equilibrium Sequences of the Roche ellipsoids (REs)
and the irrotational Roche Riemann ellipsoids (IRREs) with
$p=1$ and $\bar{a}/m_1=3$.

\vspace{0.5cm}

Table.II. Equilibrium Sequences of the REs and the IRREs with
$p=1$ and $\bar{a}/m_1=5$.

\vspace{0.5cm}

Table.III. Equilibrium Sequences of the REs and the IRREs with
$p=1$ and $\bar{a}/m_1=8$.

\vspace{0.5cm}

Table.IV. Equilibrium Sequences of the REs and the IRREs with
$p=0.1$ and $\bar{a}/m_1=3$.

\vspace{0.5cm}

Table.V. Equilibrium Sequences of the REs and the IRREs with
$p=0.1$ and $\bar{a}/m_1=5$.

\vspace{0.5cm}

Table.VI. Equilibrium Sequences of the REs and the IRREs with
$p=0.1$ and $\bar{a}/m_1=8$.

\vspace{0.5cm}

Table.VII. Comparison of our results ($\bar{a}/m_1 = 5$)
with that of Kidder, Will, \& Wiseman[7]
in the cases of $p=1$ and $0.1$.

\vspace{0.5cm}

Table.VIII. Comparison of our results ($p=1$)
with that of Lai, Rasio, \& Shapiro[18]
in the cases of $\bar{a}/m_1=5$ and $8$.

\newpage

\begin{table}
 \begin{center}
  \begin{tabular}{ccccl|ccccl}
   \multicolumn{10}{c}{$p=1$} \\ \hline
   \multicolumn{10}{c}{$\bar{a}/m_1=3$} \\ \hline
    \multicolumn{5}{c|}{Roche Sequences} &
    \multicolumn{5}{c}{Irrotational Roche-Riemann Sequences} \\ \hline
  $a_3/a_1$&$a_2/a_1$&$\tilde{\Omega^2}$&$\tilde{J}$&$R/r_s$&
  $a_3/a_1$&$a_2/a_1$&$\tilde{\Omega^2}$&$\tilde{J}$&$R/r_s$\\ \hline
   0.950 & 0.966 & 1.87(-2)     & 2.97    & 4.75 &
   0.950 & 0.948 & 2.69(-2)     & 2.87    & 4.31 \\
         &       &              &         &      &
   0.918 & 0.912 & 4.10(-2)     & 2.86    & 3.87 $\dagger$\\
   0.912 & 0.939 & 3.19(-2)     & 2.94    & 4.13 $\dagger$&
         &       &              &         &      \\
   0.900 & 0.929 & 3.59(-2)     & 2.94    & 4.00 &
   0.900 & 0.891 & 4.81(-2)     & 2.86    & 3.72 \\
   0.850 & 0.889 & 5.19(-2)     & 2.97    & 3.65 &
   0.850 & 0.833 & 6.53(-2)     & 2.88    & 3.46 \\
   0.800 & 0.847 & 6.67(-2)     & 3.01    & 3.44 &
   0.800 & 0.773 & 7.92(-2)     & 2.92    & 3.31 \\
   0.750 & 0.801 & 8.03(-2)     & 3.05    & 3.30 &
   0.750 & 0.715 & 9.02(-2)     & 2.95    & 3.22 \\
   0.700 & 0.754 & 9.23(-2)     & 3.10    & 3.21 &
   0.700 & 0.657 & 9.87(-2)     & 3.00    & 3.17 \\
   0.650 & 0.704 & 1.03(-1)     & 3.15    & 3.14 &
   0.650 & 0.601 & 1.05(-1)     & 3.04    & 3.14 \\
   0.600 & 0.652 & 1.11(-1)     & 3.20    & 3.10 &
   0.600 & 0.547 & 1.09(-1)     & 3.09    & 3.12 \\
         &       &              &         &      &
   0.577 & 0.524 & 1.11(-1)     & 3.12    & 3.12 $\ddagger$\\
   0.550 & 0.599 & 1.17(-1)     & 3.26    & 3.08 &
   0.550 & 0.496 & 1.12(-1)     & 3.15    & 3.12 \\
   0.514 & 0.560 & 1.20(-1)     & 3.30    & 3.07 $\ddagger$&
         &       &              &         &      \\
   0.500 & 0.544 & 1.21(-1)     & 3.32    & 3.07 &
   0.500 & 0.447 & 1.12(-1)     & 3.21    & 3.14 \\
   0.450 & 0.489 & 1.22(-1)     & 3.38    & 3.09 &
   0.450 & 0.400 & 1.11(-1)     & 3.28    & 3.17 \\
   0.400 & 0.433 & 1.19(-1)     & 3.45    & 3.12 &
   0.400 & 0.354 & 1.07(-1)     & 3.36    & 3.21 \\
   0.350 & 0.376 & 1.14(-1)     & 3.53    & 3.18 &
   0.350 & 0.310 & 1.02(-1)     & 3.45    & 3.28 \\
   0.300 & 0.320 & 1.05(-1)     & 3.63    & 3.27 &
   0.300 & 0.267 & 9.33(-2)     & 3.56    & 3.37 \\
   0.250 & 0.264 & 9.17(-2)     & 3.75    & 3.41 &
   0.250 & 0.225 & 8.23(-2)     & 3.69    & 3.51 \\
   0.200 & 0.209 & 7.53(-2)     & 3.90    & 3.62 &
   0.200 & 0.182 & 6.83(-2)     & 3.86    & 3.72 \\
  \end{tabular}
 \end{center}
 \caption{}
 \label{table1a}
\end{table}%

\begin{table}
 \begin{center}
  \begin{tabular}{ccccl|ccccl}
   \multicolumn{10}{c}{$p=1$} \\ \hline
   \multicolumn{10}{c}{$\bar{a}/m_1=5$} \\ \hline
    \multicolumn{5}{c|}{Roche Sequences} &
    \multicolumn{5}{c}{Irrotational Roche-Riemann Sequences} \\ \hline
   $a_3/a_1$&$a_2/a_1$&$\tilde{\Omega^2}$&$\tilde{J}$&$R/r_s$&
   $a_3/a_1$&$a_2/a_1$&$\tilde{\Omega^2}$&$\tilde{J}$&$R/r_s$\\ \hline
   0.950 & 0.967 & 1.97(-2)      & 3.30    & 7.24 &
   0.950 & 0.947 & 2.90(-2)      & 3.12    & 6.47 \\
   0.900 & 0.931 & 3.81(-2)      & 3.18    & 5.99 &
   0.900 & 0.890 & 5.25(-2)      & 3.02    & 5.48 \\
   0.850 & 0.892 & 5.56(-2)      & 3.15    & 5.40 &
   0.850 & 0.831 & 7.15(-2)      & 2.99    & 5.05 \\
         &       &               &         &      &
   0.832 & 0.810 & 7.73(-2)      & 2.99    & 4.95 $\dagger$\\
   0.828 & 0.874 & 6.29(-2)      & 3.15    & 5.22 $\dagger$&
           &         &                 &           &        \\
   0.800 & 0.850 & 7.19(-2)      & 3.15    & 5.04 &
   0.800 & 0.771 & 8.67(-2)      & 2.99    & 4.81 \\
   0.750 & 0.806 & 8.68(-2)      & 3.17    & 4.81 &
   0.750 & 0.711 & 9.85(-2)      & 3.01    & 4.66 \\
   0.700 & 0.759 & 1.00(-1)      & 3.20    & 4.64 &
   0.700 & 0.653 & 1.08(-1)      & 3.04    & 4.57 \\
   0.650 & 0.709 & 1.12(-1)      & 3.24    & 4.53 &
   0.650 & 0.597 & 1.14(-1)      & 3.08    & 4.52 \\
   0.600 & 0.657 & 1.21(-1)      & 3.28    & 4.46 &
   0.600 & 0.543 & 1.19(-1)      & 3.14    & 4.50 \\
         &       &               &         &      &
   0.582 & 0.524 & 1.20(-1)      & 3.16    & 4.50 $\ddagger$\\
   0.550 & 0.604 & 1.28(-1)      & 3.34    & 4.42 &
   0.550 & 0.492 & 1.21(-1)      & 3.20    & 4.51 \\
   0.513 & 0.564 & 1.31(-1)      & 3.38    & 4.41 $\ddagger$&
         &       &               &         &      \\
   0.500 & 0.549 & 1.32(-1)      & 3.40    & 4.41 &
   0.500 & 0.443 & 1.21(-1)      & 3.28    & 4.53 \\
   0.450 & 0.493 & 1.33(-1)      & 3.48    & 4.44 &
   0.450 & 0.396 & 1.20(-1)      & 3.36    & 4.58 \\
   0.400 & 0.436 & 1.30(-1)      & 3.57    & 4.50 &
   0.400 & 0.351 & 1.15(-1)      & 3.47    & 4.66 \\
   0.350 & 0.378 & 1.23(-1)      & 3.68    & 4.60 &
   0.350 & 0.308 & 1.09(-1)      & 3.59    & 4.77 \\
   0.300 & 0.321 & 1.13(-1)      & 3.81    & 4.75 &
   0.300 & 0.265 & 9.96(-2)      & 3.75    & 4.94 \\
   0.250 & 0.265 & 9.85(-2)      & 3.98    & 4.99 &
   0.250 & 0.223 & 8.75(-2)      & 3.94    & 5.17 \\
   0.200 & 0.209 & 8.03(-2)      & 4.22    & 5.34 &
   0.200 & 0.182 & 7.23(-2)      & 4.19    & 5.51 \\
  \end{tabular}
 \end{center}
 \caption{}
 \label{table1b}
\end{table}%

\begin{table}
 \begin{center}
  \begin{tabular}{ccccl|ccccl}
   \multicolumn{10}{c}{$p=1$} \\ \hline
   \multicolumn{10}{c}{$\bar{a}/m_1=8$} \\ \hline
    \multicolumn{5}{c|}{Roche Sequences} &
    \multicolumn{5}{c}{Irrotational Roche-Riemann Sequences} \\ \hline
   $a_3/a_1$&$a_2/a_1$&$\tilde{\Omega^2}$&$\tilde{J}$&$R/r_s$&
   $a_3/a_1$&$a_2/a_1$&$\tilde{\Omega^2}$&$\tilde{J}$&$R/r_s$\\ \hline
   0.950 & 0.968 & 2.02(-2)      & 3.82    &11.00 &
   0.950 & 0.947 & 3.04(-2)      & 3.56    & 9.71 \\
   0.900 & 0.932 & 3.95(-2)      & 3.61    & 8.98 &
   0.900 & 0.890 & 5.53(-2)      & 3.37    & 8.13 \\
   0.850 & 0.894 & 5.78(-2)      & 3.54    & 8.03 &
   0.850 & 0.830 & 7.54(-2)      & 3.30    & 7.44 \\
   0.800 & 0.853 & 7.50(-2)      & 3.51    & 7.46 &
   0.800 & 0.769 & 9.13(-2)      & 3.28    & 7.06 \\
         &       &               &         &      &
   0.792 & 0.760 & 9.35(-2)      & 3.28    & 7.02 $\dagger$\\
   0.774 & 0.830 & 8.34(-2)      & 3.51    & 7.24 $\dagger$&
         &       &               &         &      \\
   0.750 & 0.809 & 9.09(-2)      & 3.51    & 7.07 &
   0.750 & 0.709 & 1.04(-1)      & 3.29    & 6.84 \\
   0.700 & 0.762 & 1.05(-1)      & 3.53    & 6.81 &
   0.700 & 0.651 & 1.13(-1)      & 3.31    & 6.70 \\
   0.650 & 0.712 & 1.17(-1)      & 3.56    & 6.63 &
   0.650 & 0.595 & 1.20(-1)      & 3.36    & 6.62 \\
   0.600 & 0.661 & 1.27(-1)      & 3.60    & 6.51 &
   0.600 & 0.541 & 1.24(-1)      & 3.42    & 6.59 \\
         &       &               &         &      &
   0.584 & 0.524 & 1.25(-1)      & 3.44    & 6.59 $\ddagger$\\
   0.550 & 0.607 & 1.35(-1)      & 3.66    & 6.45 &
   0.550 & 0.490 & 1.27(-1)      & 3.49    & 6.60 \\
   0.513 & 0.566 & 1.38(-1)      & 3.72    & 6.43 $\ddagger$&
         &       &               &         &      \\
   0.500 & 0.552 & 1.39(-1)      & 3.74    & 6.44 &
   0.500 & 0.441 & 1.27(-1)      & 3.59    & 6.64 \\
   0.450 & 0.495 & 1.40(-1)      & 3.83    & 6.48 &
   0.450 & 0.394 & 1.25(-1)      & 3.70    & 6.72 \\
   0.400 & 0.438 & 1.37(-1)      & 3.94    & 6.57 &
   0.400 & 0.349 & 1.20(-1)      & 3.83    & 6.85 \\
   0.350 & 0.380 & 1.30(-1)      & 4.08    & 6.74 &
   0.350 & 0.306 & 1.13(-1)      & 3.99    & 7.03 \\
   0.300 & 0.322 & 1.18(-1)      & 4.26    & 6.99 &
   0.300 & 0.264 & 1.04(-1)      & 4.19    & 7.29 \\
   0.250 & 0.265 & 1.03(-1)      & 4.49    & 7.37 &
   0.250 & 0.223 & 9.06(-2)      & 4.45    & 7.66 \\
   0.200 & 0.210 & 8.33(-2)      & 4.80    & 7.94 &
   0.200 & 0.181 & 7.47(-2)      & 4.78    & 8.22 \\
  \end{tabular}
 \end{center}
 \caption{}
 \label{table1c}
\end{table}%

\begin{table}
 \begin{center}
  \begin{tabular}{ccccl|ccccl}
   \multicolumn{10}{c}{$p=0.1$} \\ \hline
   \multicolumn{10}{c}{$\bar{a}/m_1=3$} \\ \hline
    \multicolumn{5}{c|}{Roche Sequences} &
    \multicolumn{5}{c}{Irrotational Roche-Riemann Sequences} \\ \hline
   $a_3/a_1$&$a_2/a_1$&$\tilde{\Omega^2}$&$\tilde{J}$&$R/r_s$&
   $a_3/a_1$&$a_2/a_1$&$\tilde{\Omega^2}$&$\tilde{J}$&$R/r_s$\\ \hline
         &       &               &         &      &
   0.992 & 0.992 & 2.42(-3)      & 2.70    & 3.26 $\dagger$\\
   0.990 & 0.992 & 2.39(-3)      & 2.71    & 3.27 $\dagger$&
         &       &               &         &      \\
   0.950 & 0.959 & 9.94(-3)      & 2.85    & 2.35 &
   0.950 & 0.949 & 1.17(-2)      & 2.89    & 2.27 \\
   0.900 & 0.914 & 1.80(-2)      & 3.02    & 2.08 &
   0.900 & 0.896 & 2.04(-2)      & 3.06    & 2.03 \\
   0.850 & 0.868 & 2.51(-2)      & 3.16    & 1.96 &
   0.850 & 0.843 & 2.76(-2)      & 3.19    & 1.93 \\
   0.800 & 0.821 & 3.12(-2)      & 3.26    & 1.88 &
   0.800 & 0.789 & 3.35(-2)      & 3.29    & 1.86 \\
   0.750 & 0.772 & 3.65(-2)      & 3.35    & 1.84 &
   0.750 & 0.735 & 3.84(-2)      & 3.37    & 1.82 \\
   0.700 & 0.722 & 4.10(-2)      & 3.42    & 1.80 &
   0.700 & 0.682 & 4.23(-2)      & 3.43    & 1.79 \\
   0.650 & 0.672 & 4.47(-2)      & 3.48    & 1.78 &
   0.650 & 0.629 & 4.53(-2)      & 3.48    & 1.77 \\
   0.600 & 0.621 & 4.75(-2)      & 3.52    & 1.76 &
   0.600 & 0.577 & 4.74(-2)      & 3.51    & 1.76 \\
   0.550 & 0.569 & 4.93(-2)      & 3.56    & 1.75 &
   0.550 & 0.526 & 4.86(-2)      & 3.54    & 1.76 \\
         &       &               &         &      &
   0.525 & 0.501 & 4.89(-2)      & 3.55    & 1.76 $\ddagger$\\
   0.500 & 0.517 & 5.02(-2)      & 3.58    & 1.75 $\ddagger$&
   0.500 & 0.476 & 4.90(-2)      & 3.55    & 1.76 \\
   0.450 & 0.465 & 5.01(-2)      & 3.59    & 1.75 &
   0.450 & 0.427 & 4.85(-2)      & 3.56    & 1.76 \\
   0.400 & 0.412 & 4.89(-2)      & 3.58    & 1.76 &
   0.400 & 0.379 & 4.70(-2)      & 3.55    & 1.78 \\
   0.350 & 0.360 & 4.66(-2)      & 3.57    & 1.78 &
   0.350 & 0.331 & 4.46(-2)      & 3.54    & 1.79 \\
   0.300 & 0.308 & 4.29(-2)      & 3.54    & 1.81 &
   0.300 & 0.284 & 4.11(-2)      & 3.51    & 1.82 \\
   0.250 & 0.255 & 3.80(-2)      & 3.49    & 1.85 &
   0.250 & 0.238 & 3.64(-2)      & 3.46    & 1.86 \\
   0.200 & 0.204 & 3.17(-2)      & 3.42    & 1.92 &
   0.200 & 0.192 & 3.05(-2)      & 3.40    & 1.93 \\
  \end{tabular}
 \end{center}
 \caption{}
 \label{table2a}
\end{table}%

\begin{table}
 \begin{center}
  \begin{tabular}{ccccl|ccccl}
   \multicolumn{10}{c}{$p=0.1$} \\ \hline
   \multicolumn{10}{c}{$\bar{a}/m_1=5$} \\ \hline
    \multicolumn{5}{c|}{Roche Sequences} &
    \multicolumn{5}{c}{Irrotational Roche-Riemann Sequences} \\ \hline
   $a_3/a_1$&$a_2/a_1$&$\tilde{\Omega^2}$&$\tilde{J}$&$R/r_s$&
   $a_3/a_1$&$a_2/a_1$&$\tilde{\Omega^2}$&$\tilde{J}$&$R/r_s$\\ \hline
           &         &                 &           &        &
   0.962 & 0.961 & 1.10(-2)      & 2.71    & 3.28 $\dagger$\\
   0.953 & 0.962 & 1.07(-2)      & 2.71    & 3.30 $\dagger$&
           &         &                 &           &        \\
   0.950 & 0.960 & 1.14(-2)      & 2.71    & 3.25 &
   0.950 & 0.949 & 1.38(-2)      & 2.71    & 3.09 \\
   0.900 & 0.917 & 2.12(-2)      & 2.75    & 2.79 &
   0.900 & 0.896 & 2.47(-2)      & 2.75    & 2.70 \\
   0.850 & 0.872 & 3.00(-2)      & 2.79    & 2.58 &
   0.850 & 0.841 & 3.38(-2)      & 2.80    & 2.51 \\
   0.800 & 0.825 & 3.78(-2)      & 2.83    & 2.45 &
   0.800 & 0.787 & 4.14(-2)      & 2.84    & 2.41 \\
   0.750 & 0.777 & 4.47(-2)      & 2.87    & 2.37 &
   0.750 & 0.732 & 4.76(-2)      & 2.88    & 2.34 \\
   0.700 & 0.728 & 5.06(-2)      & 2.91    & 2.31 &
   0.700 & 0.678 & 5.17(-2)      & 2.90    & 2.30 \\
   0.650 & 0.677 & 5.55(-2)      & 2.94    & 2.27 &
   0.650 & 0.624 & 5.64(-2)      & 2.93    & 2.26 \\
   0.600 & 0.626 & 5.92(-2)      & 2.97    & 2.24 &
   0.600 & 0.572 & 5.90(-2)      & 2.95    & 2.24 \\
   0.550 & 0.574 & 6.18(-2)      & 2.99    & 2.23 &
   0.550 & 0.521 & 6.05(-2)      & 2.97    & 2.24 \\
         &       &               &         &      &
   0.528 & 0.498 & 6.08(-2)      & 2.98    & 2.24 $\ddagger$\\
   0.500 & 0.521 & 6.30(-2)      & 3.00    & 2.22 &
   0.500 & 0.471 & 6.09(-2)      & 2.98    & 2.24 \\
   0.497 & 0.518 & 6.31(-2)      & 3.00    & 2.22 $\ddagger$&
         &       &               &         &      \\
   0.450 & 0.469 & 6.29(-2)      & 3.02    & 2.22 &
   0.450 & 0.422 & 6.02(-2)      & 2.99    & 2.25 \\
   0.400 & 0.416 & 6.13(-2)      & 3.02    & 2.24 &
   0.400 & 0.374 & 5.82(-2)      & 3.00    & 2.27 \\
   0.350 & 0.362 & 5.82(-2)      & 3.03    & 2.27 &
   0.350 & 0.327 & 5.50(-2)      & 3.00    & 2.30 \\
   0.300 & 0.309 & 5.34(-2)      & 3.02    & 2.32 &
   0.300 & 0.281 & 5.04(-2)      & 3.00    & 2.35 \\
   0.250 & 0.257 & 4.69(-2)      & 3.02    & 2.39 &
   0.250 & 0.236 & 4.44(-2)      & 3.00    & 2.42 \\
   0.200 & 0.204 & 3.88(-2)      & 3.01    & 2.47 &
   0.200 & 0.190 & 3.69(-2)      & 3.00    & 2.53 \\
  \end{tabular}
 \end{center}
 \caption{}
 \label{table2b}
\end{table}%

\begin{table}
 \begin{center}
  \begin{tabular}{ccccl|ccccl}
   \multicolumn{10}{c}{$p=0.1$} \\ \hline
   \multicolumn{10}{c}{$\bar{a}/m_1=8$} \\ \hline
    \multicolumn{5}{c|}{Roche Sequences} &
    \multicolumn{5}{c}{Irrotational Roche-Riemann Sequences} \\ \hline
   $a_3/a_1$&$a_2/a_1$&$\tilde{\Omega^2}$&$\tilde{J}$&$R/r_s$&
   $a_3/a_1$&$a_2/a_1$&$\tilde{\Omega^2}$&$\tilde{J}$&$R/r_s$\\ \hline
   0.950 & 0.961 & 1.23(-2)      & 2.82    & 4.61 &
   0.950 & 0.949 & 1.53(-2)      & 2.78    & 4.34 \\
   0.900 & 0.919 & 2.34(-2)      & 2.75    & 3.88 &
   0.900 & 0.895 & 2.80(-2)      & 2.73    & 3.70 \\
   0.850 & 0.874 & 3.35(-2)      & 2.73    & 3.53 &
   0.850 & 0.840 & 3.86(-2)      & 2.72    & 3.41 \\
         &       &               &         &       &
   0.849 & 0.840 & 3.87(-2)      & 2.72    & 3.41 $\dagger$\\
   0.832 & 0.858 & 3.69(-2)      & 2.73    & 3.45 $\dagger$&
         &       &               &         &      \\
   0.800 & 0.828 & 4.27(-2)      & 2.74    & 3.32 &
   0.800 & 0.785 & 4.76(-2)      & 2.72    & 3.24 \\
   0.750 & 0.781 & 5.09(-2)      & 2.74    & 3.19 &
   0.750 & 0.729 & 5.49(-2)      & 2.73    & 3.13 \\
   0.700 & 0.732 & 5.80(-2)      & 2.75    & 3.09 &
   0.700 & 0.674 & 6.07(-2)      & 2.74    & 3.06 \\
   0.650 & 0.681 & 6.39(-2)      & 2.77    & 3.02 &
   0.650 & 0.620 & 6.51(-2)      & 2.75    & 3.01 \\
   0.600 & 0.630 & 6.85(-2)      & 2.78    & 2.97 &
   0.600 & 0.568 & 6.82(-2)      & 2.77    & 2.98 \\
   0.550 & 0.578 & 7.17(-2)      & 2.79    & 2.95 &
   0.550 & 0.516 & 6.99(-2)      & 2.78    & 2.97 \\
         &       &               &         &      &
   0.529 & 0.496 & 7.02(-2)      & 2.78    & 2.97 $\ddagger$\\
   0.500 & 0.525 & 7.33(-2)      & 2.81    & 2.94 &
   0.500 & 0.466 & 7.03(-2)      & 2.79    & 2.97 \\
   0.494 & 0.519 & 7.33(-2)      & 2.81    & 2.94 $\ddagger$&
         &       &               &         &      \\
   0.450 & 0.472 & 7.32(-2)      & 2.82    & 2.94 &
   0.450 & 0.418 & 6.93(-2)      & 2.81    & 2.98 \\
   0.400 & 0.418 & 7.13(-2)      & 2.84    & 2.97 &
   0.400 & 0.371 & 6.69(-2)      & 2.82    & 3.02 \\
   0.350 & 0.364 & 6.75(-2)      & 2.85    & 3.02 &
   0.350 & 0.324 & 6.31(-2)      & 2.84    & 3.07 \\
   0.300 & 0.311 & 6.17(-2)      & 2.87    & 3.09 &
   0.300 & 0.279 & 5.76(-2)      & 2.86    & 3.14 \\
   0.250 & 0.258 & 5.39(-2)      & 2.90    & 3.21 &
   0.250 & 0.234 & 5.05(-2)      & 2.89    & 3.26 \\
   0.200 & 0.205 & 4.41(-2)      & 2.93    & 3.38 &
   0.200 & 0.189 & 4.17(-2)      & 2.93    & 3.44 \\
  \end{tabular}
 \end{center}
 \caption{}
 \label{table2c}
\end{table}%

\newpage

\begin{table}
 \begin{center}
  \begin{tabular}{|c|c|c|c|}
                &$p$                       & 0.1 &  1  \\ \hline \hline
   [7]'s results&$r_{ISCO}/M_{tot}$        & 6.5 & 7.0 \\ \hline
   Our results  &$R^{RE}_{ISCO}/M_{tot}$   & 6.6 &10.4 \\ \hline
   Our results  &$R^{IRRE}_{ISCO}/M_{tot}$ & 6.6 & 9.9
  \end{tabular}
 \end{center}
 \caption{}
\end{table}%

\vspace{5cm}

\begin{table}
 \begin{center}
  \begin{tabular}{|c|c|c|c|}
   \multicolumn{4}{c}{$p=1$}            \\ \hline
                 &$\bar{a}/m_1$            &  5   &  8 \\ \hline \hline
   [18]'s results&$r_m/M_{tot}$            &  8.5 & 12  \\ \hline
   Our results   &$R^{RE}_{ISCO}/M_{tot}$  & 10.4 & 14.5 \\ \hline
   Our results   &$R^{IRRE}_{ISCO}/M_{tot}$&  9.9 & 14.0
  \end{tabular}
 \end{center}
 \caption{}
\end{table}%

\newpage

\begin{center}
  {\large FIGURE CAPTIONS}
\end{center}

\vspace{0.5cm}

Fig.1(a). The effective potentials of a test particle 
in the Schwarzschild metric (left)
and the pseudo-Newtonian potential (right) as
a function of the normalized distance $R/r_s$.
The vertical axes denote
$\Psi_{Sch} = \sqrt{\left( 1-\frac{2M}{R} \right)
\left( \frac{J_{Sch}^2}{R^2} +1 \right)}-1$
and $\Psi_{pseudo} = -\frac{M}{R-r_s} +
\frac{J_{pseudo}^2}{2R^2}$, respectively.
The dots express the place of circular orbits.
The value of the effective potential of the Schwarzschild black hole
at the infinity is shifted to zero.

\vspace{0.5cm}

Fig.1(b). The fractional deviation of the circular orbits of
the pseudo-Newtonian potential from those of
the Schwarzschild effective potential as a function of the
angular momentum. Circular orbits with
different angular momentum related
as $J_{pseudo} = ( \frac{9}{8})^{1/2} J_{Sch}$
are compared (see text).

\vspace{0.5cm}

Fig.2(a). The total angular momentum
$\tilde{J} = J_{tot}/\{m_1 m_2 (r_s/M_{tot})^{1/2}\}$
of the equilibrium sequence as a function of $R/r_s$
in the case that the mass ratio is $p=1$.
Thin and thick lines express
the REs and the IRREs cases, respectively.
Solid, dotted, and dashed lines denote $\bar{a}/m_1=$
$3$, $5$ and $8$, respectively.

\vspace{0.5cm}

Fig.2(b). The axial ratios $a_2/a_1$ and $a_3/a_1$ of
the equilibrium sequence as a function of $R/r_s$
in the case of the mass ratio $p=1$.
The top figure represents $a_2/a_1$, and
the bottom does $a_3/a_1$.
The conventions are the same as in Fig.2(a).

\vspace{0.5cm}

Fig.3(a). The total angular momentum $\tilde{J}$
of the equilibrium sequence as a function of $R/r_s$
in the case of $p=0.1$.
The region around the minimum of the lines is magnified
and shown in the upper right corner.
The conventions are the same as in Fig.2(a).

\vspace{0.5cm}

Fig.3(b). The axial ratios $a_2/a_1$ and $a_3/a_1$ of
the equilibrium sequence as a function of $R/r_s$
in the case of $p=0.1$.
The top figure represents $a_2/a_1$, and
the bottom does $a_3/a_1$.
The conventions are the same as in Fig.2(a).

\vspace{0.5cm}

Fig.4. The relation between the mean radius of the primary
$\bar{a}/m_1$ and the separation of the binary $R_{ISCO}/r_s$
at the ISCO in the case of $p=1$.
Thin lines denote the case of the Newtonian potential
as an interaction potential, and thick lines
denote the case of the generalized pseudo-Newtonian potential.
The solid and dotted lines represent
the IRREs and the REs, respectively.

\vspace{0.5cm}

Fig.5 The relation between the mean radius of the primary
$\bar{a}/m_1$ and the separation of the binary $R_{ISCO}/r_s$
at the ISCO in the case of $p=0.1$.
The conventions are the same as in Fig.4.

\newpage

\begin{figure}[ht]
  \vspace{1cm}
  \centerline{\epsfysize 15cm \epsfxsize 15cm \epsfbox{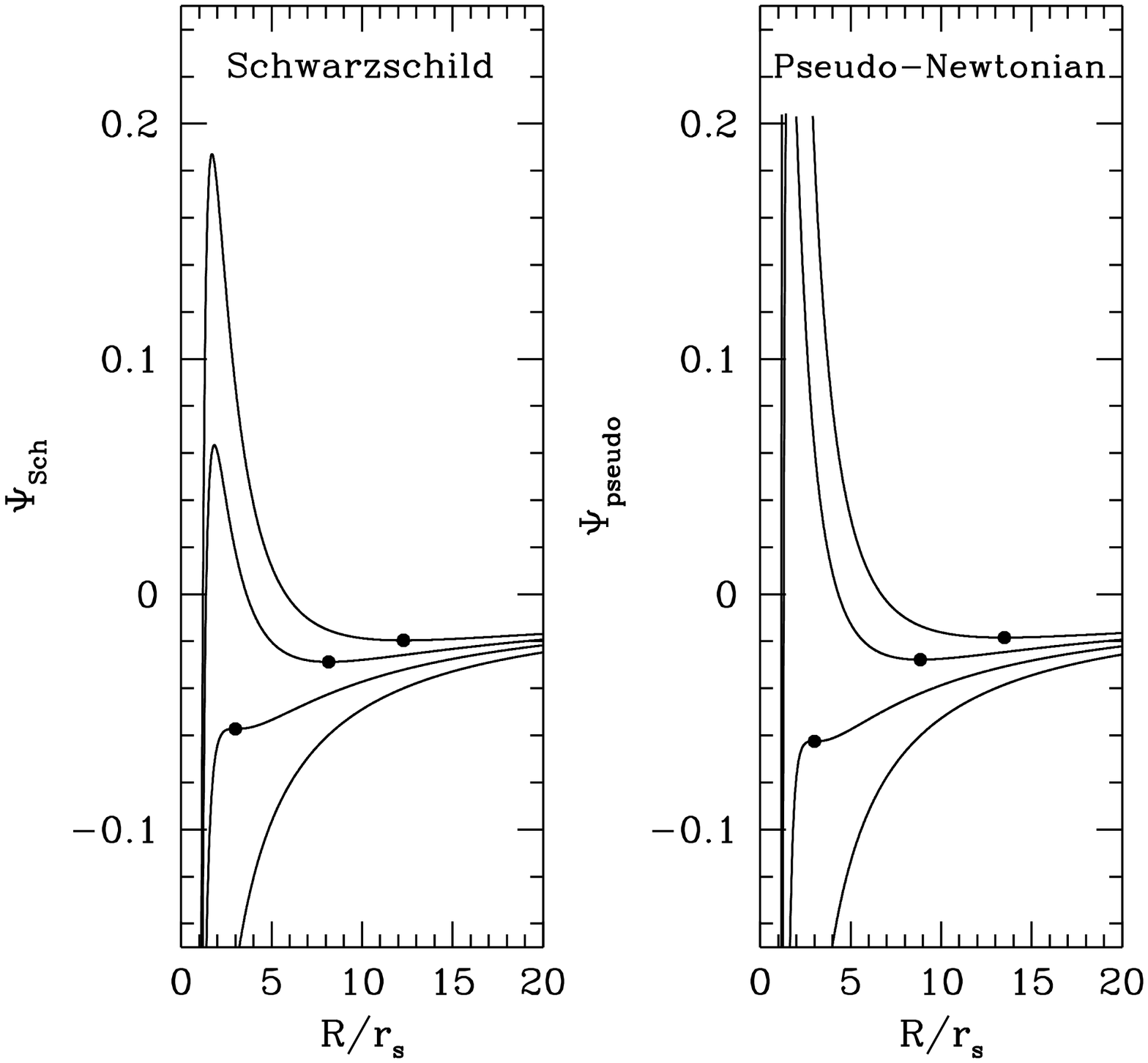}}
  \vspace{0.5cm}
  \label{effect}
\end{figure}%

\vspace{3cm}


\hspace{250pt}
Fig.1(a)

\newpage

\begin{figure}[ht]
  \vspace{1cm}
  \centerline{\epsfysize 13cm \epsfxsize 15cm \epsfbox{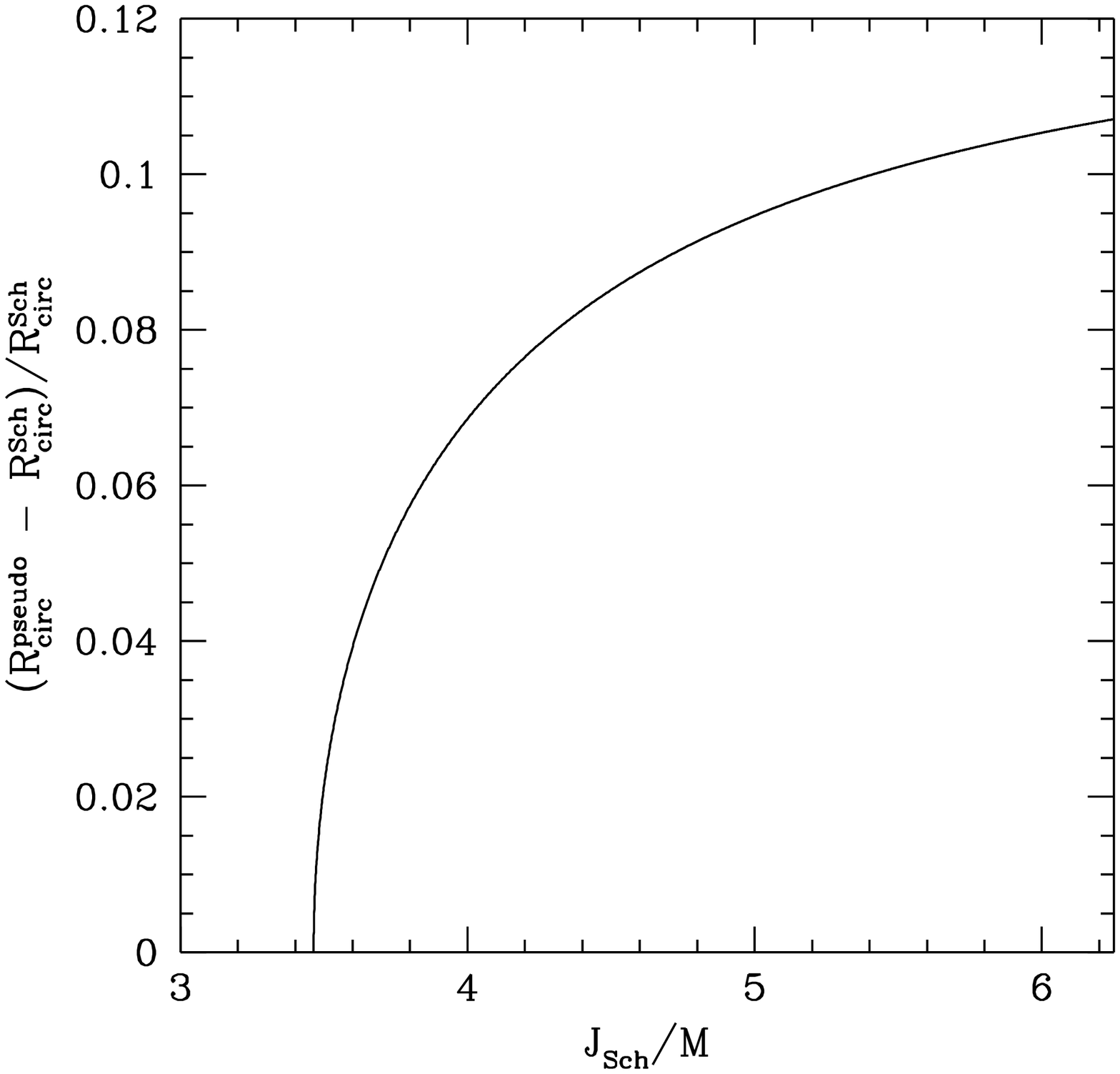}}
  \vspace{0.5cm}
  \label{orbit}
\end{figure}%

\vspace{6cm}


\hspace{250pt}
Fig.1(b)

\newpage

\begin{figure}[ht]
  \vspace{1cm}
  \centerline{\epsfysize 13cm \epsfxsize 15cm \epsfbox{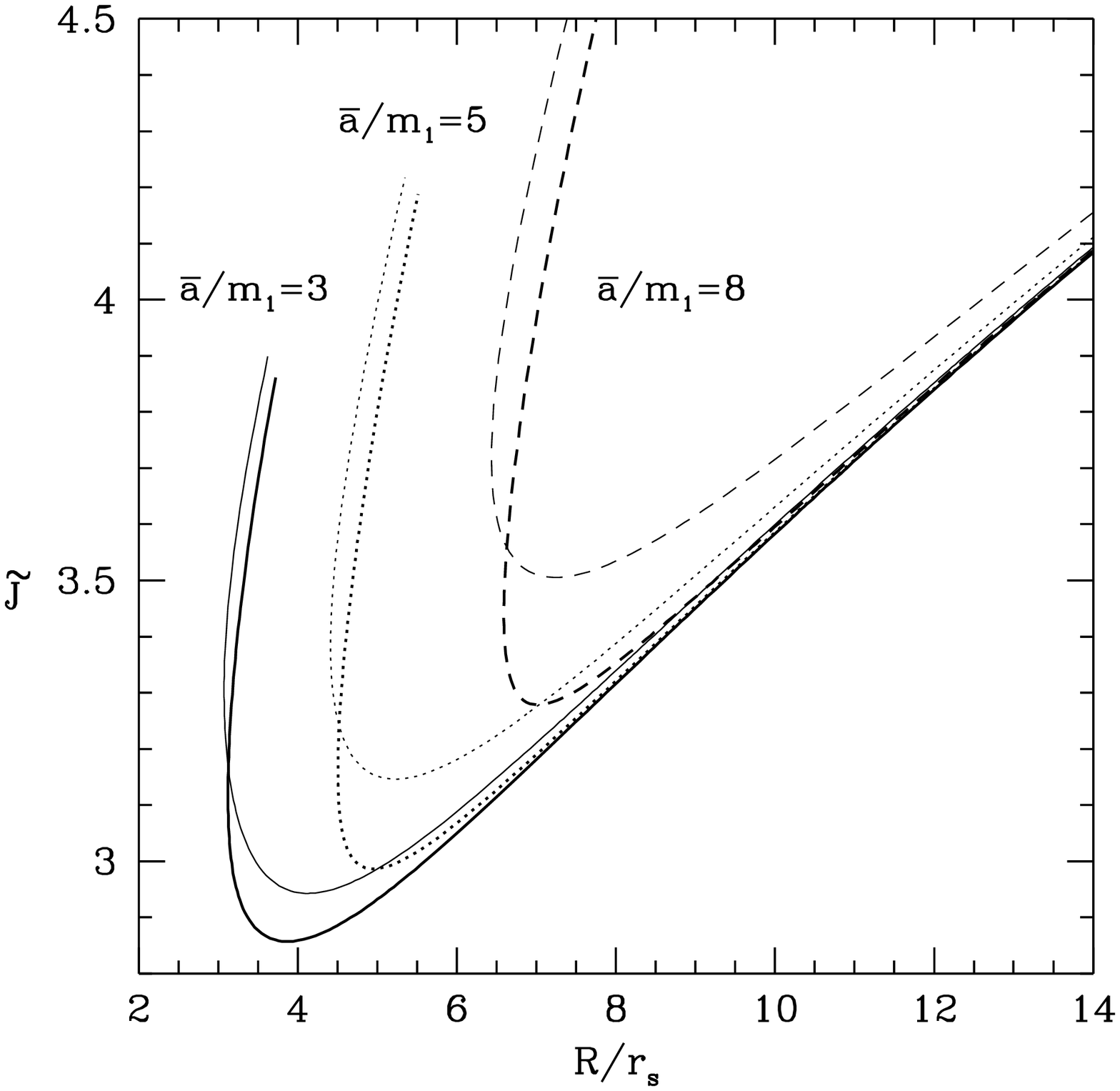}}
  \vspace{0.5cm}
  \label{jp1}
\end{figure}%

\vspace{6cm}


\hspace{250pt}
Fig.2(a)

\newpage

\begin{figure}[ht]
  \vspace{1cm}
  \centerline{\epsfysize 15cm \epsfxsize 15cm \epsfbox{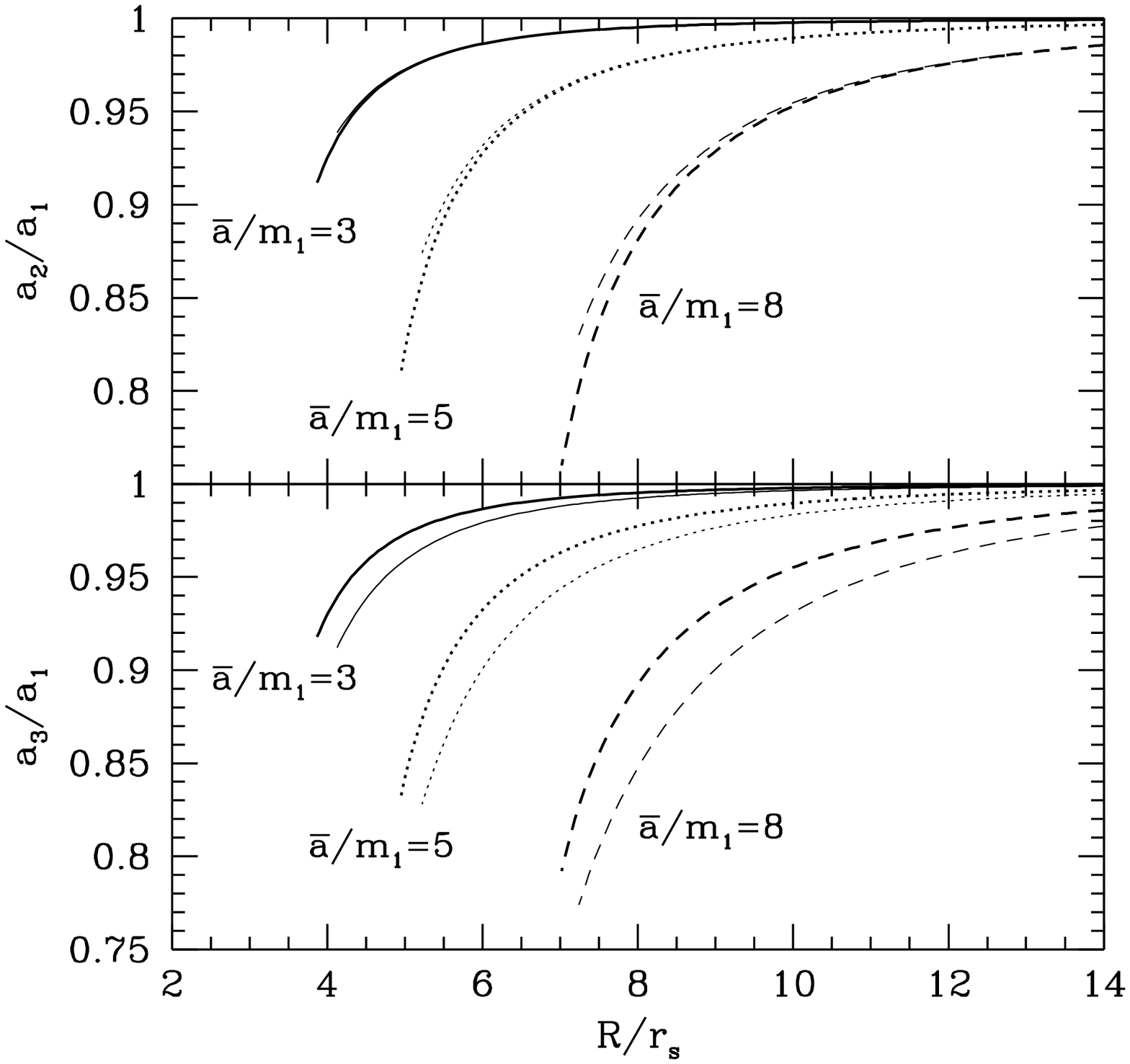}}
  \vspace{0.5cm}
  \label{stp1}
\end{figure}%

\vspace{4cm}


\hspace{250pt}
Fig.2(b)

\newpage

\begin{figure}[ht]
  \vspace{1cm}
  \centerline{\epsfysize 13cm \epsfxsize 15cm \epsfbox{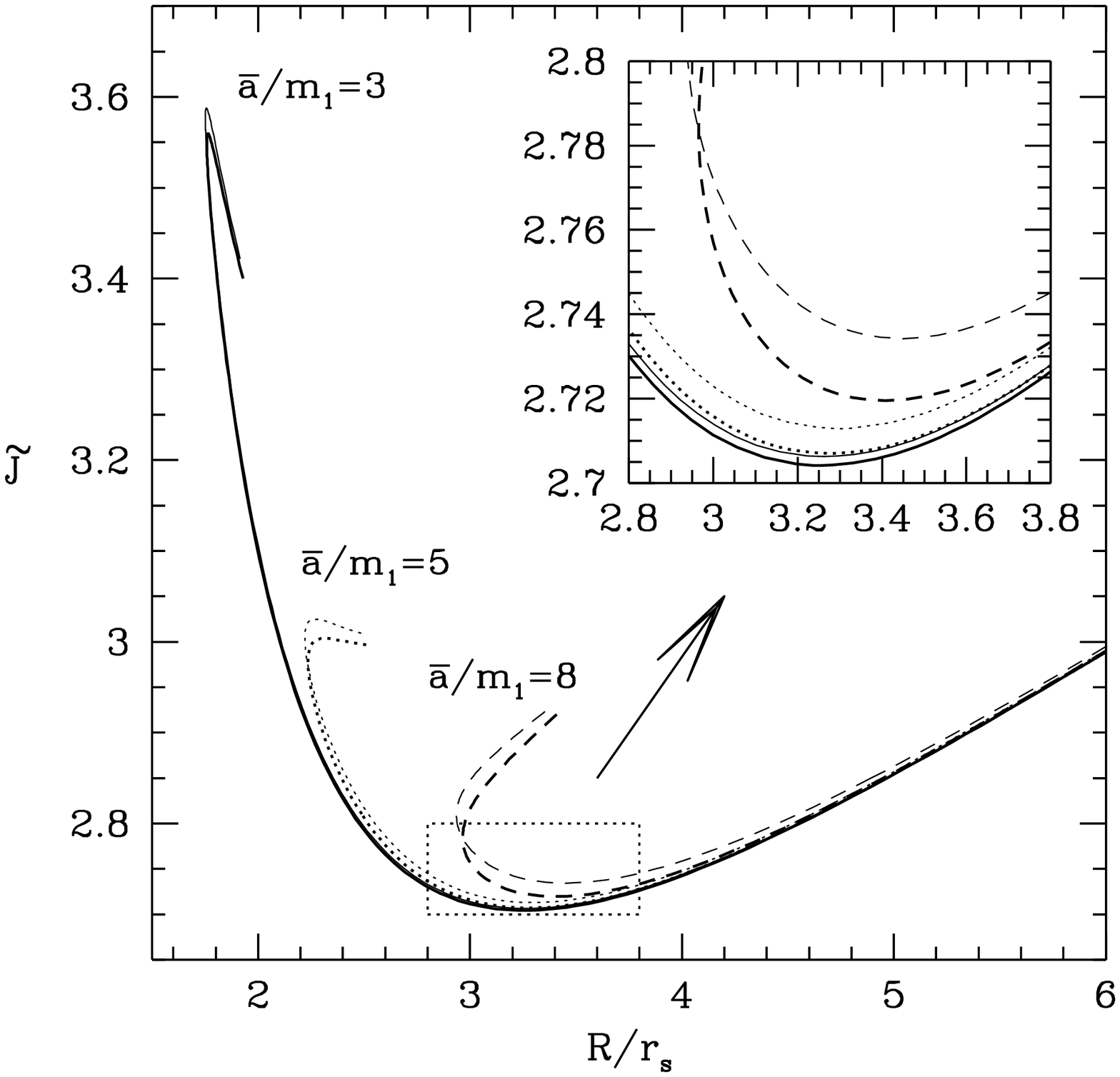}}
  \vspace{0.5cm}
  \label{jp01}
\end{figure}%

\vspace{6cm}


\hspace{250pt}
Fig.3(a)

\newpage

\begin{figure}[ht]
  \vspace{1cm}
  \centerline{\epsfysize 15cm \epsfxsize 15cm \epsfbox{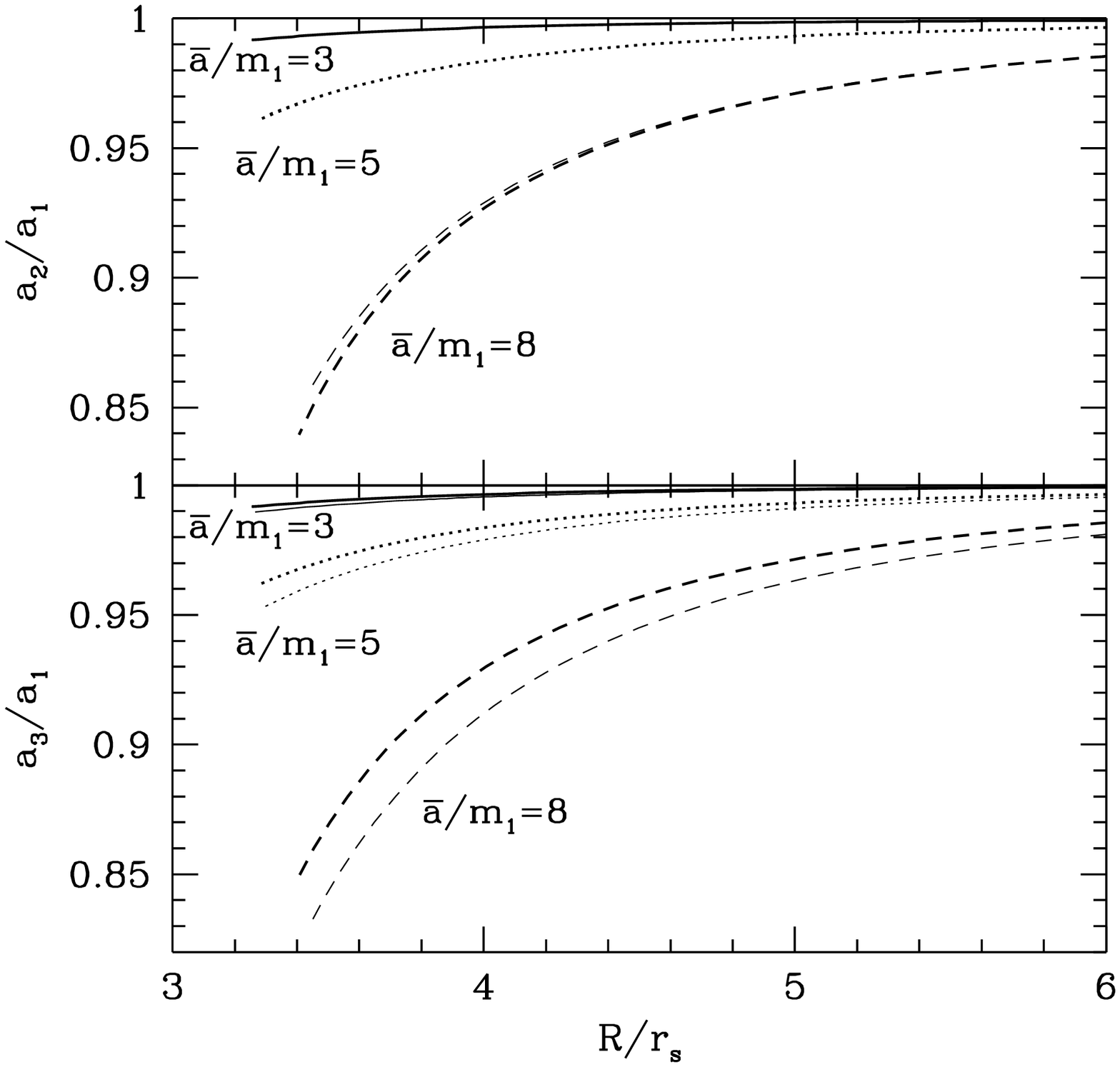}}
  \vspace{0.5cm}
  \label{stp01}
\end{figure}%

\vspace{4cm}


\hspace{250pt}
Fig.3(b)

\newpage

\begin{figure}[htb]
  \vspace{1cm}
  \centerline{\epsfysize 13cm \epsfxsize 15cm \epsfbox{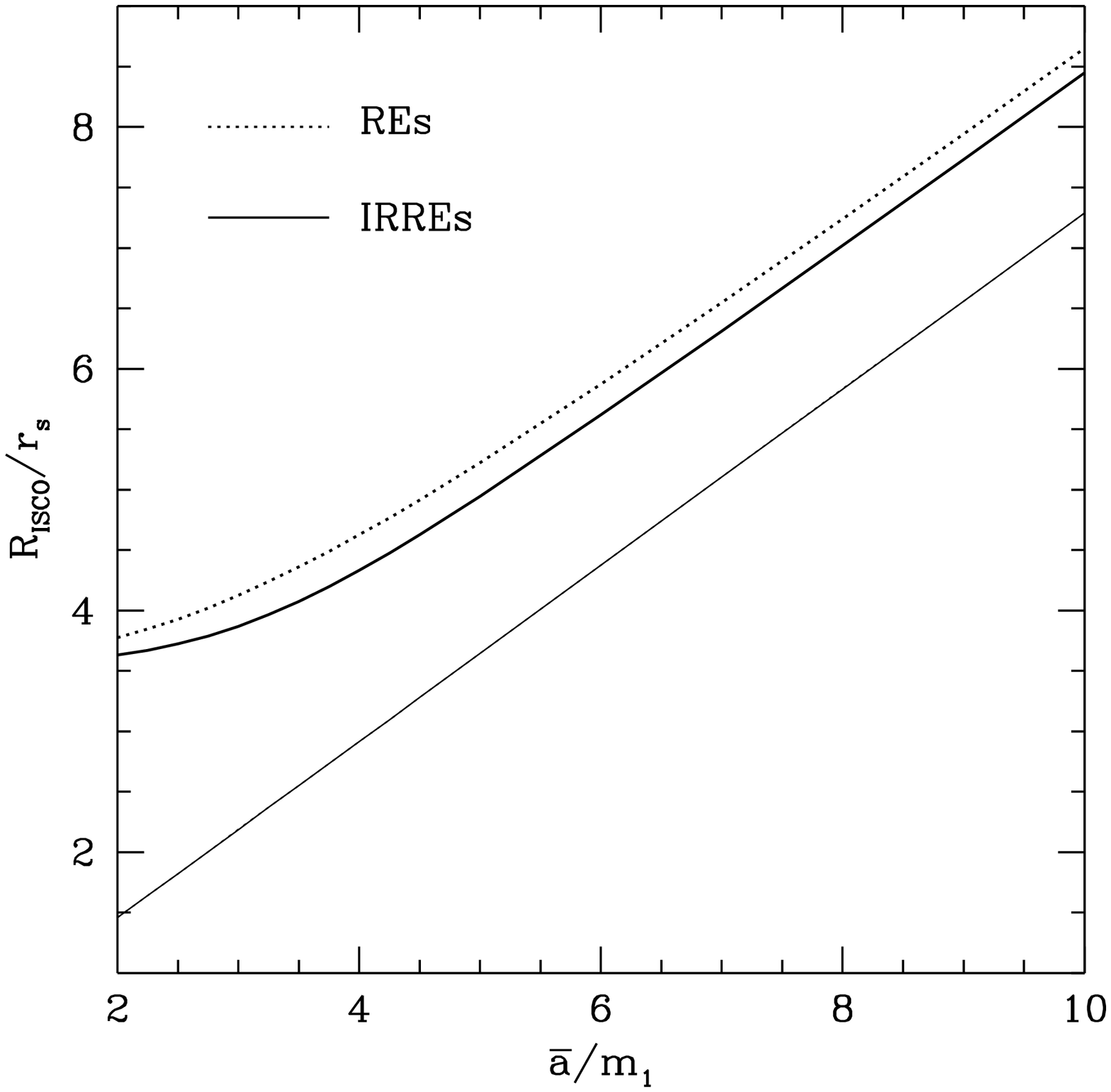}}
  \vspace{0.5cm}
  \label{jap1}
\end{figure}%

\vspace{6cm}


\hspace{250pt}
Fig.4

\newpage

\begin{figure}[htb]
  \vspace{1cm}
  \centerline{\epsfysize 13cm \epsfxsize 15cm \epsfbox{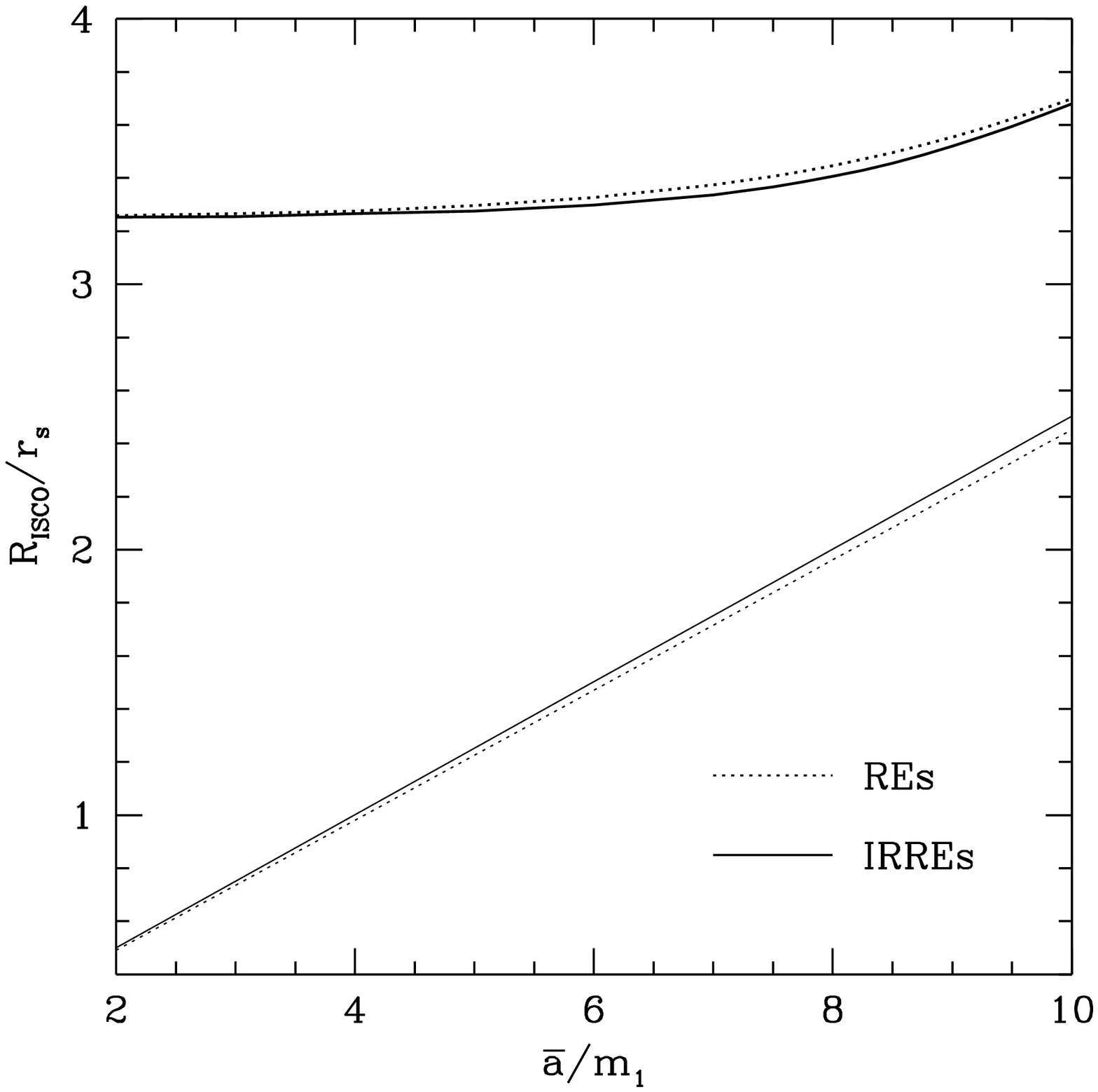}}
  \vspace{0.5cm}
  \label{jap01}
\end{figure}%

\vspace{6cm}


\hspace{250pt}
Fig.5

\end{document}